\documentclass[%
 reprint,
 amsmath,amssymb,
 aps,
 prb,
]{revtex4-2}

\usepackage{graphicx}% Include figure files
\usepackage{dcolumn}% Align table columns on decimal point
\usepackage{bm}% bold math

\usepackage{amsmath}
\usepackage{amssymb}
\usepackage{color,soul}
\usepackage{ulem}  %ky for sout
\usepackage{amsfonts}%
\usepackage[caption=false]{subfig}
\usepackage{comment}
\usepackage[dvipsnames]{xcolor}
\usepackage[capitalize]{cleveref}
\usepackage{float}
\usepackage{tabularx} 
\usepackage{array}

\newcommand\sto{SrTiO$_3$}
\newcommand\bto{BaTiO$_3$}

\begin{document}
\preprint{APS/123-QED}
\title{Dislocation-induced flexoelectricity  \\
in SrTiO$_3$ nanostructure 
from first principles}% Force line breaks with \\
\author{Kunihiko Yamauchi}
\email{yamauchi.kunihiko.es@osaka-u.ac.jp}
\author{Thi Phuong Thao Nguyen}
\author{Tamio Oguchi}
\affiliation{%
 Center for Spintronics Research Network, Osaka University, 
1-3 Machikaneyama, Toyonaka, Osaka 560-8531, Japan
}%
\date{\today}% It is always \today, today,
\begin{abstract}
%We study flexoelectric polarization induced by misfit dislocation by means of first-principles calculation.A nano-size \sto\ atomic structure was created in a supercell model, in which we found that the oxygen ionic displacement near misfit-dislocation cores causes the sizable polarization. We also discuss the electronic state and lattice strain effect on polarization. 

Flexoelectricity refers to a linear coupling between the electric polarization and the strain gradient, such as bending or asymmetric compression. 
This effect is enhanced in nano-scale structures, where grain boundaries or dislocation cores induce the strain gradient. 
In this study, we theoretically investigate the flexoelectric 
polarization induced by misfit dislocations in a thin film. A nano-scale dislocation structure is modeled in a periodic \sto\ supercell, and then the structure is optimized by using neural-network-potential and first-principles approaches. We point out that a pyramidal TiO$_5$ coordination forms near the dislocation cores, which in turn dominantly causes the sizable flexoelectric polarization. 
%The flexoelectric polarization is evaluated based on a point-charge model. This reveals the sizable polarization is mainly caused by oxygen ionic displacement of the TiO$_5$ pyramids. 

%the effects of the electronic state and lattice strain on polarization.
\end{abstract}

\maketitle
%\tableofcontents

\section{\label{sec:intro}Introduction}
Flexoelectricity is a property of dielectric materials where electric polarization is induced by a strain gradient due to inhomogeneous deformation.\cite{Zubko2013}  
This flexoelectric coupling between polarization and strain gradient is distinct from the piezoelectric coupling between polarization and homogeneous strain. 
Such macroscopic strain gradients can be induced by bending \cite{Nguyen2013, Stengel2013} or by applying asymmetric compression \cite{Cross2006} in materials. A stronger flexoelectric effect has %attracted increasing attention because the strain-gradient scale is inverse to the material dimensions. Nano-scale flexoelectricity has 
been observed in nano-scale structures such as thin films \cite{Lee2011} and in nanowires \cite{Liu2012}. %We hypothesized that the strain gradient between coherently strained and relaxed cubic phases could show flexo-electric polarization10,11,14 if the strain and film thicknesses were adjusted appropriately. 
Recently, 
nano-scale  flexoelectricity has been engineered by exploiting grain boundaries \cite{Wu2022} and dislocations \cite{Yamahara2021}.  
Due to the broken translational symmetry at dislocations, a strain gradient naturally occurs around the dislocation cores,  significantly enhancing the electric polarization. 
%Yamahara {\it et al.}  did..
In this context, atomic-scale measurements have been performed to investigate the polarization %around dislocation cores 
in a grain boundary of \sto\ bicrystal \cite{Gao2018} and at crack tips at SrMnO$_3$/SrTiO$_3$ interface \cite{Wang2020}; 
these strain fields %resulting from these dislocations 
indeed induce the flexoelectric polarization. 
%and sizable polarization of $\sim$ 28 $\mu$C/cm$^2$ was found. %at distances smaller than 1 nm away from the dislocations.

%Hereinafter, we study 
%On the other hand, the 
{\it ''Misfit dislocations"} are caused by mechanical misfit strain at the interface. 
When the material is grown on the substrate whose lattice constants are slightly smaller than the material,  misfit dislocation forms to relax the lattice strain as shown in Fig. \ref{fig:model} (a). 
Experimental measurements have reported that in BaTiO$_3$ thin films grown on (001) \sto\ substrate with 2\%  lattice mismatch, misfit dislocations contribute to the strain relaxation and set a critical thickness above which the film retains its bulk structure \cite{Sun2004, Suzuki1999}. The presence of dislocation cores leads to cation non-stoichiometry and strong structural distortion at the oxide interface, which can locally enhance electric polarization via the flexoelectric effect \cite{arrendondo2010, Yamahara2021}. Although a scanning transmission electron microscopy measurement provides the atomic resolution, oxygen ions are not visible due to their weak scattering cross-section. Nevertheless, the oxygen ionic positions are crucial for evaluating electric polarization, as discussed in \bto\ and other ferroelectric transition-metal oxides.\cite{cohen1997, ederer2005, yamauchi2008, okuyama2020} 

In this study,  we perform a first-principles density-functional-theory (DFT) simulation to illuminate the microscopic mechanism of the dislocation-induced polarization in a transition-metal oxide. 
We selected \sto\ as our model system since it is an incipient ferroelectric oxide that can cause significant electric polarization sensitive to applied strain while its crystal structure is centrosymmetric at the ground state.\cite{zhong1995, zhong1996, Zubko2007} 
Our investigation will concentrate on elucidating how ionic displacements around the misfit-dislocation core can lead to spontaneous electric polarization, with a focus on the position of oxygen ions. 
%\ky{Enrich the motivation of this study!}

\section{ STRUCTURAL model AND COMPUTATIONAL DETAILS}

\begin{figure}[htb]
\begin{center}
\includegraphics[width=7cm]{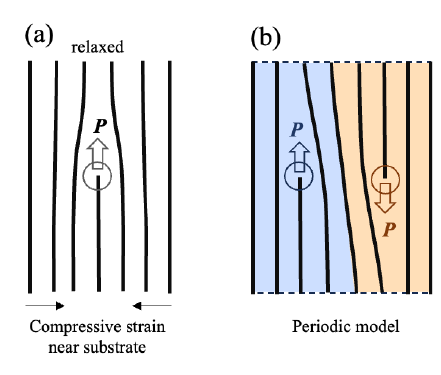}
\caption{\label{fig:model} 
(a) A schematic illustration of a misfit dislocation near a substrate/film interface. Thick lines denote the alignment of ions. A circle indicates the dislocation core, and a block arrow indicates electric polarization induced by flexoelectricity. (b) Its periodic structural model with two counteracting domains painted in blue and orange colors. 
}
\end{center}
\end{figure}

\begin{figure*}[htb]
\begin{center}
\includegraphics[width=15cm]{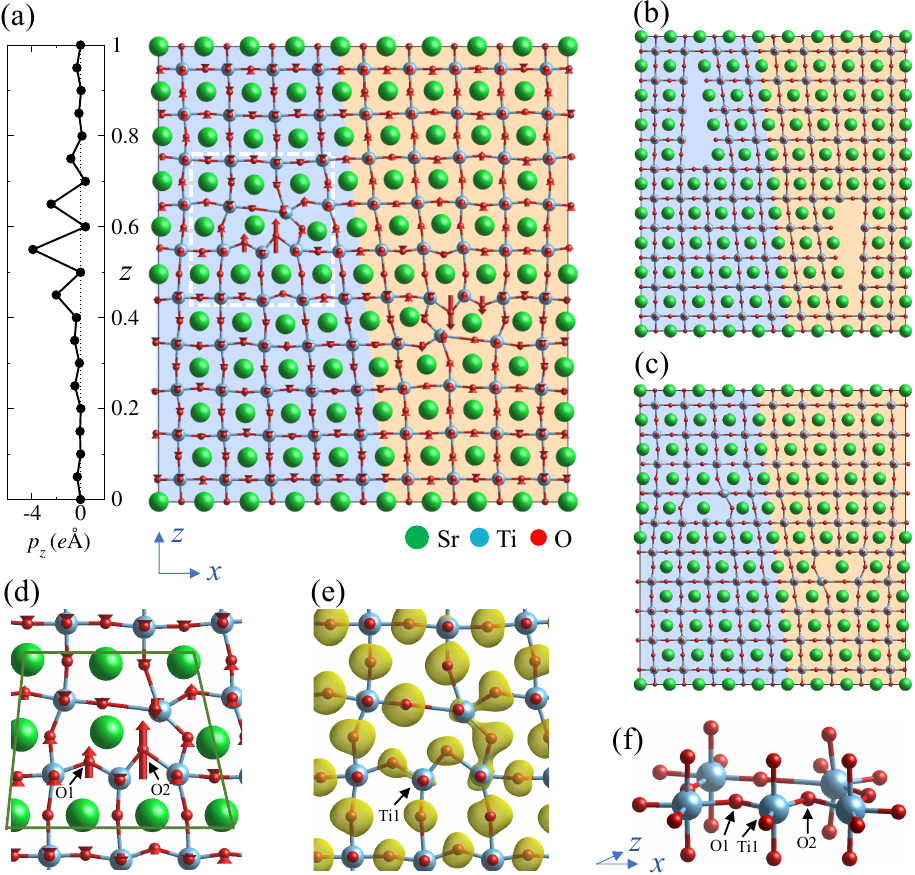}
\caption{\label{fig:widefig} 
(a) Optimized atomic structure of the misfit-dislocation model. Purple and orange areas show two counteracting polar domains.  
The ionic displacement along the $z$ direction is shown by red arrows. The Ti-O atomic bonds are drawn to display the pristine \sto\ structure as a guide to the eye. 
An inset graph shows the local dipole moment $p_z$ (the unit is a product of elementary charge $e$ and \AA) by summing up the ionic contributions in $0\leq x<0.5$ range.  
(b) Initial atomic structure. 
(c) Non-polar reference structure to evaluate the dipole moment and the polarization. 
(d) Ionic displacement near the dislocation core. A green line indicates a region to calculate the local polarization. 
(e) Partial charge density around the valence band minimum (VBM) in the energy range of $E_{\rm VBM}$-2.0 eV $<E<E_{\rm VBM}$ in the TiO$_2$ plane. 
(f) 3D atomic structure around the dislocation core. 
}
\end{center}
\end{figure*}

To perform DFT calculations, we designed a periodic structure of the misfit dislocation in \sto\ supercell as schematically shown in Fig.\ref{fig:model} (b).  
This superstructure consists of two polar domains ({\it i.e.}, left and right domains), and their net spontaneous polarization cancels out with the inversion symmetry. 
The \sto\ bulk cubic structure with the experimental lattice constants ($a$ = $b$ = $c$ = 3.905 \AA) was used as a block unit to build the dislocation structure.
%To simulate the dislocation in a computational process, 
First, we built a  10$\times$1$\times$10 supercell. And then apporximately eight percent of atoms (9 Sr, 8 Ti, and 25 O atoms) 
were removed and lattice constant $a$ was shortened by 10 \% along the $x$ axis as shown in Fig.\ref{fig:widefig} (b). 
By doing so, the middle layer ($z$=0.5) containing 11 cells feels 10 \% of compressive strain while the top and bottom layers ($z$ = 0.0 or 1.0) containing 10 cells are nearly in equilibrium. 
The intermediate layers ($0<z<0.5$ and $0.5<z<1.0$) are going to be reconstructed to reduce the energy in the optimization process.  
The supercell contains 91 Sr, 92 Ti, and 275 O atoms in the end;  this satisfies the charge neutrality with their nominal charges, {\it i.e.}, $91\times(+2) + 92\times(+4) + 275\times(-2) = 0$.

%\begin{figure}[htb]
%\begin{center}
%\includegraphics[width=6cm]{structure.pdf}
%\caption{\label{fig:structure} 
%(a) Initial and (b) optimized structure for the dislocation \sto. Purple and orange areas show two counteracting polar domains.  
%}
%\end{center}
%\end{figure}

To save the computational time to optimize the large atomic structure with 458 atoms, a pre-optimization step was performed using a graph neural network-based machine-learning interatomic potential implemented in CHGNet \cite{chgnet} code. 
After that, atomic positions were relaxed again until the maximum force was less than 0.01 eV \AA$^{-1}$ using the Vienna Ab initio Simulation Package (VASP)\cite{vasp1,vasp2}. 
During the atomic optimization, 
the lattice constants and  positions of the Sr and O atoms located at $z$=0.0 and 0.5 layers were fixed;  
since the dislocations arise from the lattice mismatch between substrate and film, the two layers at $z=0$ and $z=0.5$ represent the two limits, one away from the interface and one near the interface, respectively. The structure keeps the centrosymmetric $C_{2h}$ symmetry  %during the optimization 
so that the net electric polarization stays zero while the local dipole moments are induced by the strain gradient caused by the dislocation. %We found that the structure optimized by CHGNet code provides an almost identical structure to that optimized by VASP code; the change in the atomic position is around 0.\red{??}\AA\ at maximum.  
%For the GB, we found that the charge-neutral system is gapless.  
Electronic structure calculations were performed using the VASP code. The projected augmented wave (PAW) potential was used to address the interactions between ions and electrons. The exchange-correlation potential was treated by the Perdew–Burke–Ernzerhof functional \cite{pbe}. A plane-wave cutoff of 600 eV and a 
Monkhorst-Pack $k$-point sampling with 1$\times$2$\times$1 mesh were used. 
The Born effective charges were computed by using density functional perturbation theory \cite{dfpt}.

\section{Results and Discussions}

Figure \ref{fig:widefig} (a) shows the final optimized structure projected on the $xz$ plane. It is found that the preoptimized structure (not shown) by the CHGNet code is slightly different from the one fully optimized by the VASP code; the change in the atomic position is around 0.36 \AA\ at maximum.  
During the structural optimization, Ti-O bonds were reconstructed around the dislocation cores so as to remove the dangling bonds.

\begin{figure}[htb]
\begin{center}
\includegraphics[width=6cm]{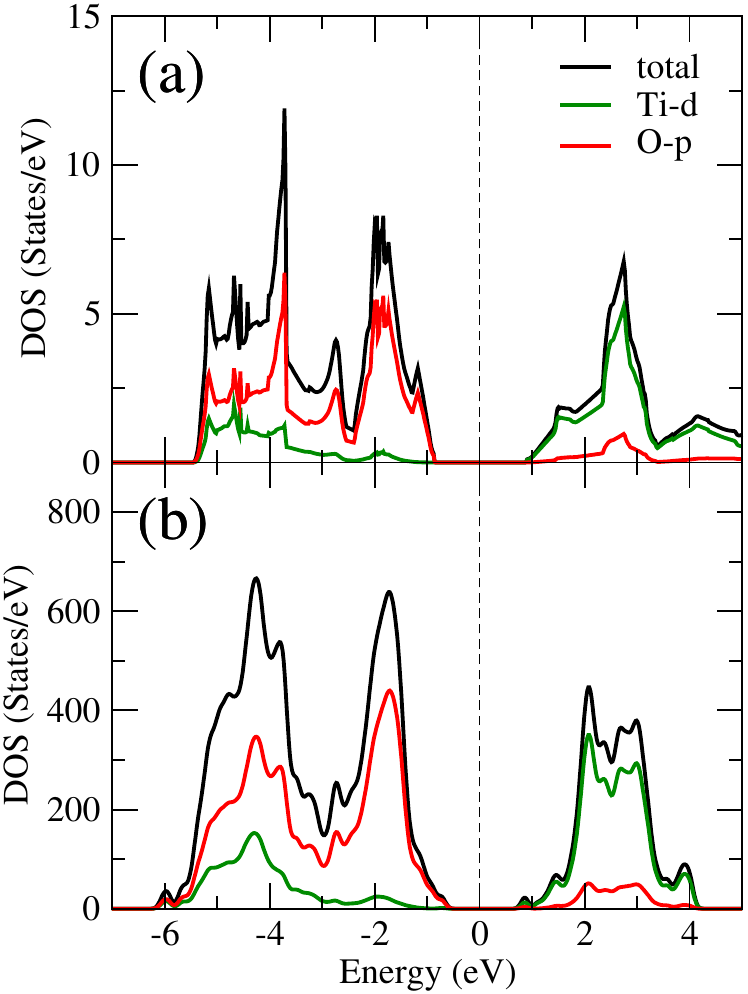}
\caption{\label{fig:pdos} 
Total DOS and projected DOS of Ti 3$d$ state and O 2$p$ state in (a) bulk perovskite \sto\ and (b) the \sto\ dislocation structure. Contributions from the same atomic kinds were summed up. 
}
\end{center}
\end{figure}

As shown in Fig.\ref{fig:pdos}, the density of states (DOS) shows the insulating electronic state with the wide band gap $E_{\rm g}$ = 1.55 eV, slightly smaller than the calculated value for bulk \sto\ as $E_{\rm g}$ = 1.95 eV. 
The DOS shows that the atomic structure does not have any dangling bonds that often cause deep impurity levels inside the band gap.

Figure \ref{fig:widefig} (d) shows the $z$ component of the ionic displacement  
near the dislocation core. 
We label the O ions with the largest ionic displacement O1 and O2.  
Ti1 atoms and surrounding O atoms including O1 and O2 atoms form a five-coordinated trigonal bipyramid because the one O atom is missing from the original octahedral coordinate. 
%and the Ti1-O1 and Ti1-O2 ionic bonds show the strong $pd$ hybridization as shown in Fig.\ref{fig:widefig} (d-e). 
This geometry leads to the large ionic displacements along the $z$ axis as 0.74 \AA\ and 1.36 \AA\ for O1 and O2, respectively, with respect to the non-polar reference structure shown in Fig.\ref{fig:model} (c).  
Recalling the fact that the ionic displacement in ferroelectric oxides such as BaTiO$_3$ is typically 0.1 \AA\ order, the dislocation-induced displacement is tremendously large. 

%the spontaneous polarization induced by the two oxygen displacements in three \sto\ unit cells was roughly estimated as 37.3$\mu$C/cm$^2$.

\begin{figure}[htb]
\begin{center}
\includegraphics[width=6cm]{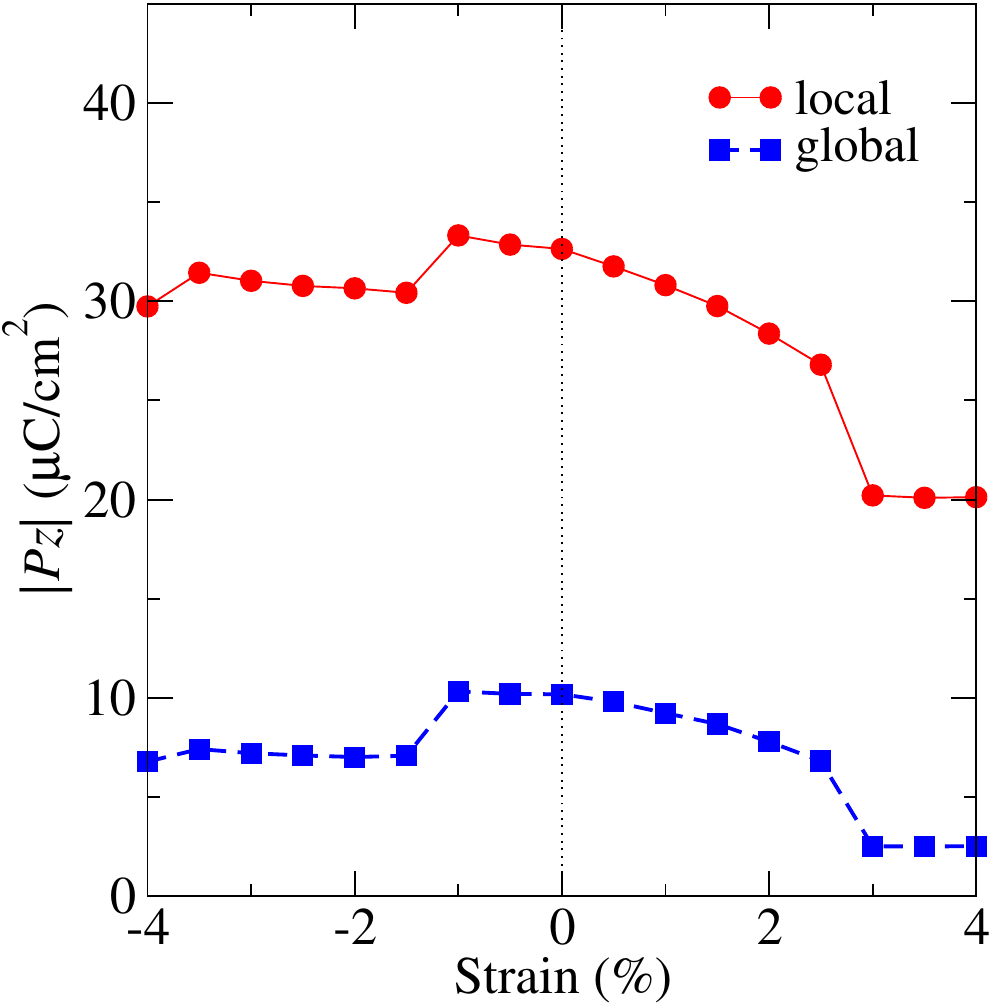}
\caption{\label{fig:pion} 
The flexoelectric polarization $|P_z|$ evaluated in a local region near the dislocation core and in a global region (left half of the supercell). 
The trend of $|P_z|$ under the $a$-axis lattice strain (positive: tensile; negative: compressive) is shown. 
To reduce the computational cost, the atomic structure is optimized only by CHGNet code for this result. 
}
\end{center}
\end{figure}

If we focus on the left-half domain of the supercell, the structure is polar under the non-uniform strain, continuously changing the size from  4$a_{0}$
at top to  5$a_{0}$ at bottom, where $a_{0}$ is the lattice constant of bulk \sto. 
In this case, the polarization may be considered as a response to an applied strain and a macroscopic strain gradient: 
\begin{equation} \label{eq:polarization}
P_i = e_{ijk} \epsilon_{jk} + \mu_{ijkl} \frac{\partial \epsilon_{jk}}{\partial x_l}, 
\end{equation}
where $P$ is electric polarization, $\epsilon$ strain, and $x$ position. 
While $e$ is a 3rd-order tensor called piezoelectric $e$-constant, $\mu$ is a 4th-order tensor called flexoelectric constant. 
Assuming a rough cancellation of the piezoelectric effect between the upper part under tensile strain and the lower part under compressive strain, we can neglect the first term. As shown in Fig.\ref{fig:pion}, the piezoelectric response of the polarization $P_{z}$ is rather small in a region of -2\% $< \epsilon_{xx} < $ 2\%. 
%Indeed, applied $x$-axis strain does not affect the polarization significantly . 
This is because the polarization is mainly caused by the TiO$_5$ bipyramidal coordinate and its bond framework stays robust in the range of the strain. 

The electric polarization can be qualitatively evaluated by summing up the dipole moments. 
Based on the point-charge model, the dipole moment is defined as the following:
\begin{equation}\label{eq:dipole}
    \vec{p}=\sum_a q_a \vec{r_a}, 
\end{equation}
where $q_a$ is the elementary charge for $a$-th ion; for simplicity, we use the nominal valence: +2, +4, and -2 for Sr, Ti, and O ion, and $\vec{r_a}$ is the ionic displacement from the non-polar reference structure, in which the $z$ component of all the ions is set same as the initial structure as shown in Fig.\ref{fig:widefig} (c). 
Then the polarization is calculated by dividing the total dipole moment, $\vec{p}$, by the volume of the system, $V$ :
\begin{equation}\label{eq:pcm}
    \vec{P}=\frac{\vec{p}}{V} = \frac{1}{V}\sum_a q_a \vec{r_a}. 
\end{equation}
We note that the polarization and the related flexoelectric effect are extensive properties; hence we need to define a region to evaluate the polarization per dislocation core. 
If the net polarization is evaluated in the supercell, it becomes zero due to the centrosymmetric atomic structure. 
A specific region (a green-lined trapezoid) is defined near the dislocation core as shown in Fig.\ref{fig:widefig} (d) and its volume was roughly estimated as 4.5 units of \sto\ perovskite cell: $V_{\rm l}$=($3.905$\AA$)^{3}\times 4.5=267.96$ \AA$^3$. 
As shown in Fig.\ref{fig:widefig} (a), the dipole moment is strongly enhanced in this region. 
A more global region is also defined as the supercell's left half (0$\le x <0.5$). 
Figure \ref{fig:pion} shows the calculated polarization in the local and the global regions. 
The local polarization is evaluated as $P_z^{\rm l}=$32.6 $\mu$C/cm$^2$,  whereas the polarization in the global region is $P_z^{\rm g} = $10.2 $\mu$C/cm$^2$. 
To evaluate more realistic polarization, we can replace the nominal charge $q_a$ in Eq.\ref{eq:pcm} by the Born effective charge tensor, calculated by VASP code, 
\begin{equation}
Z^*_{a,ij}=V\frac{\partial P_{i}}{\partial r_{a,j}}. 
\end{equation}
This value represents a practical way to characterize the anisotropic nature of the $pd$ hybridization along Ti–O bonds \cite{zhong.BEC.1994, yamauchi2013}. 
For example, $Z^*$ tensors of O1 and O2 ions,   
\begin{equation}
Z^*_{\rm O1} = \begin{pmatrix}
-4.7 & 0.0 & 0.1 \\
 0.0 & -1.6 & 0.0 \\
 0.7 & 0.0 & -1.5 \\
 \end{pmatrix},  
 \\
Z^*_{\rm O2} = \begin{pmatrix}
-3.5 & 0.0 & -1.2 \\
 0.0 & -1.6 & 0.0 \\
 1.0 & 0.0 & -3.4 \\
\end{pmatrix},  
\end{equation}
show the strong Ti1-O1 bonding along the $x$ direction and the strong Ti1-O2 bonding along the $x$ and $z$ direction (cfr. Fig.\ref{fig:widefig} (e)). 
By using the $zz$ component of the BEC tensors and $r_z$, the calculated polarization $P_z^{\rm l}$ is enhanced from 32.6 to 47.0 $\mu$C/cm$^2$, pointing out the importance of the electronic contribution.  
This polarization value is larger than that of bulk \bto\ ($P \sim$27 $\mu$C/cm$^2$\cite{zhong1994, kingsmith1994}). 

In the trapezoidal region shown in Fig.\ref{fig:widefig}(d), the lattice constant $a$ (bond length between Sr atoms on average) is 3.96 \AA\ and 3.51 \AA\ at top and bottom side, respectively. 
The strain gradient is roughly estimated as $d\epsilon_{11}/dz=3.96/3.51/(3.905\times 2) =0.14$ \AA$^{-1}$. 
Finally, we evaluate the flexoelectric constant, $\mu_{3113}= -47.0/0.14 = 3.26 \times 10^{-10} $C/m per dislocation core. 
This value is comparable to the experimental coefficients in \sto, where $\mu \geq$ 3$\times10^{-10}$C/m and the flexoelectricity is similarly induced by dislocation cores in grain boundaries \cite{Gao2018}.  
This makes a contrast to the larger flexoelectricity observed in a SrMnO$_3$/\sto\ heterostructure and bulk \sto, which shows $\mu$ on the order of 10$^{-9}$C/m \cite{Wang2020, zubko2007}, where the increased polarization is induced by macroscopic cracking or bending.

%-Born effective charge?
%-visualize partial charge density? 
%-ELF?
%-Stress the mechanism to enhance the local flexoelectricity: Ti-O bonding yields a large polarization. 

\section{Summary} \label{sec:summary}

We performed the nano-scale simulation of the misfit-dislocation core generated in \sto\ and observed that the Ti and O ions form the bipyramidal coordination, resulting in significant polar ionic displacement and sizable flexoelectric polarization. 
This highlights the importance of the oxygen ionic position, which is not directly observable by scanning transmission electron microscopy. 
Our simulation approach could be instrumental in fine tuning the dislocation-induced flexoelectricity, potentially extending to other materials such as spinels and garnets in future studies.  
We hope that our finding of the polar ionic displacement caused by the misfit dislocation will be validated by forthcoming experiments.

\begin{acknowledgments}
We are grateful to Hitoshi Tabata and Munetoshi Seki for the fruitful discussions. 
This work was supported  by JST-CREST (Grant No. JPMJCR22O2) and by
Institute for Open and Transdisciplinary Research Initiatives (OTRI), Osaka University.
The computation in this work has been done using the facilities of the Supercomputer Center, the Institute for Solid State Physics, the University of Tokyo and Supercomputing System MASAMUNE-IMR in the Center for Computational Materials Science, Institute for Materials Research, Tohoku University (Project No. 202312-RDKGE-0058). The crystallographic figure was generated using the VESTA program~\cite{vesta3}.
\end{acknowledgments}

\nocite{*}

% bibtex is made by BibDesk at mac. 
%\bibliography{ref}
%\bibliographystyle{apsrev4-2} 

\end{document}